\documentclass[twocolumn]{aastex631}
\usepackage{graphicx}
\usepackage{amssymb}
\usepackage{amsmath}
\usepackage{subfigure}
\usepackage{multirow}
\usepackage{threeparttable}
\usepackage{array}
\usepackage{natbib}
\usepackage{xcolor}
\usepackage{subfigure}
\usepackage{hyperref}
\usepackage[version=4]{mhchem}

\hypersetup{
    colorlinks=true,
    linkcolor=red,
    filecolor=magenta,
    urlcolor=cyan,
}

\def\kms{~\rm km~s^{-1}}

\defcitealias{Henley:2012aa}{HS12}

\begin{document}
\title{The Solar-Cycle Temporal Variation of the Solar Wind Charge Exchange X-ray Lines}

\author{Zhijie Qu}
\affiliation{Department of Astronomy \& Astrophysics, The University of Chicago, Chicago, IL 60637, USA}
\affiliation{Department of Astronomy, University of Michigan, Ann Arbor, MI 48109, USA}

\author{Dimitra Koutroumpa}
\affiliation{LATMOS-IPSL, CNRS, UVSQ Paris-Saclay, Sorbonne Universit\'e, Guyancourt, France}

\author{Joel N. Bregman}
\affiliation{Department of Astronomy, University of Michigan, Ann Arbor, MI 48109, USA}

\author{Kip D. Kuntz}
\affiliation{The Henry A. Rowland Department of Physics and Astronomy, Johns Hopkins University, Baltimore, MD 21218, USA}
\affiliation{NASA Goddard Space Flight Center, Greenbelt, MD 20771, USA}

\author{Philip Kaaret}
\affiliation{Department of Physics and Astronomy, University of Iowa, Van Allen Hall, Iowa City, IA 52242, USA}

\correspondingauthor{Zhijie Qu}
\email{quzhijie@uchicago.edu}

\begin{abstract}
Solar wind charge exchange (SWCX) is the primary contamination to soft X-ray emission lines from the Milky Way (MW) hot gas.
We report a solar-cycle ($\approx 10$ yr) temporal variation of observed \ion{O}{7} and \ion{O}{8} emission line measurements in the {\it XMM-Newton} archive, which is tightly correlated with the solar cycle traced by the sunspot number (SSN).
This temporal variation is expected to be associated with the heliospheric SWCX.
Another observed correlation is that higher solar wind (SW) fluxes lead to higher \ion{O}{7} or \ion{O}{8} fluxes, which is due to the magnetospheric SWCX.
We construct an empirical model to reproduce the observed correlation between the line measurements and the solar activity (i.e., the SW flux and the SSN).
With this model we discovered a lag of $0.91_{-0.22}^{+0.20}$ yr between the \ion{O}{7} flux and the SSN.
This time lag is a combination of the SW transit time within the heliosphere, the lag of the neutral gas distribution responding to solar activity, and the intrinsic lag between the SSN and the launch of a high-energy SW (i.e., $\rm O^{7+}$ and $\rm O^{8+}$).
MW \ion{O}{7} and \ion{O}{8} fluxes have mean values of 5.4 L.U. and 1.7 L.U., which are reduced by $50\%$ and $30\%$, compared to studies where the SWCX contamination is not removed.
This correction also changes the determination of the density distribution and the temperature profile of the MW hot gas.
\end{abstract}

\section{Introduction}
The soft X-ray emission traces the hot gas with $T \approx 10^6$ K, which is predicted to dominate the baryons in the local Universe \citep{Cen:1999aa, Bregman:2007aa}.
Previous studies revealed that the Milky Way (MW) hosts a hot gaseous halo with $T \approx 2\times10^6$ K \citep[e.g., ][]{Henley:2010aa, Kaaret:2020aa}.
However, a long-term issue of MW X-ray studies is the foreground contamination due to the solar wind charge exchange (SWCX), which is induced when the solar wind (SW) interacts with the neutral gas within the heliosphere.
The SWCX is identified as an efficient mechanism of soft X-ray line emission \citep{Cravens:1997aa} and contaminates measurements of the \ion{O}{7} and \ion{O}{8} fluxes of the MW hot gas \citep{Henley:2013aa}.

There are two types of SWCX distinguished by different origins of the neutral gas interacting with the SW (see \citealt{Kuntz:2019aa} for a review).
First, the neutral gas in the Earth's atmosphere could interact with the SW, which is known as the magnetospheric SWCX, because it happens in the magnetosphere of the Earth.
This emission was discovered as the long-term enhancements (LTEs; variation over days) by {\it ROSAT} in the soft X-ray band \citep{Cravens:2001aa}.
Because this SWCX happens near the Earth, there is no obvious lag between the temporal variation of the local SW flux around the Earth and the observed soft X-ray enhancement.

Another origin of the neutral gas is the interplanetary medium (IPM) within the heliosphere, which leads to the heliospheric SWCX.
Using deep X-ray observations, \citet{Koutroumpa:2007aa} found that the \ion{O}{7} (\ion{O}{8}) heliospheric SWCX has a large variation from $0.3 - 4.6$ ($0.02-2.1$) L.U. (line units; counts $\rm s^{-1}~cm^{-2}~sr^{-1}$).
\citet[][hereafter \citetalias{Henley:2012aa}]{Henley:2012aa} found that the SWCX (mainly heliospheric) \ion{O}{7} and \ion{O}{8} fluxes are $\lesssim 5$ L.U. and $\lesssim 2$ L.U. using multiple epochs of observations of the same sight line in the {\it XMM-Newton} archive.

However, the temporal behavior of the heliospheric SWCX is still uncertain.
The temporal variation depends upon the distribution of the neutral gas, the propagation of the solar wind through the heliosphere, and the way that the solar wind modifies the distribution of neutral gas.
Some studies have performed detailed calculations of the SWCX on a per observation basis and used those estimates to subtract the SWCX contamination from the Galactic \ion{O}{7} and \ion{O}{8} emission (e.g. \citealt{Kaaret:2020aa}).
However, this was not performed for the widely used \ion{O}{7} and \ion{O}{8} line survey of \citepalias{Henley:2012aa} (i.e., the current largest and highest spatial resolution sample for the MW and SWCX studies).
In this letter, we investigate both types of SWCX empirically using the \ion{O}{7} and \ion{O}{8} line survey \citepalias{Henley:2012aa}, and report evidence for a long-term temporal variation of the SWCX.
We constructed an empirical model to reproduce the observed correlation between \ion{O}{7} and \ion{O}{8} line measurements and solar activity (e.g., the SW flux measured around the Earth and the sunspot number; SSN).
This model provides a sample of Galactic \ion{O}{7} and \ion{O}{8} emission that has been far better cleaned of SWCX contamination than previous studies (using the \citepalias{Henley:2012aa} survey).

\begin{figure*}
\begin{center}
\includegraphics[width=0.45\textwidth]{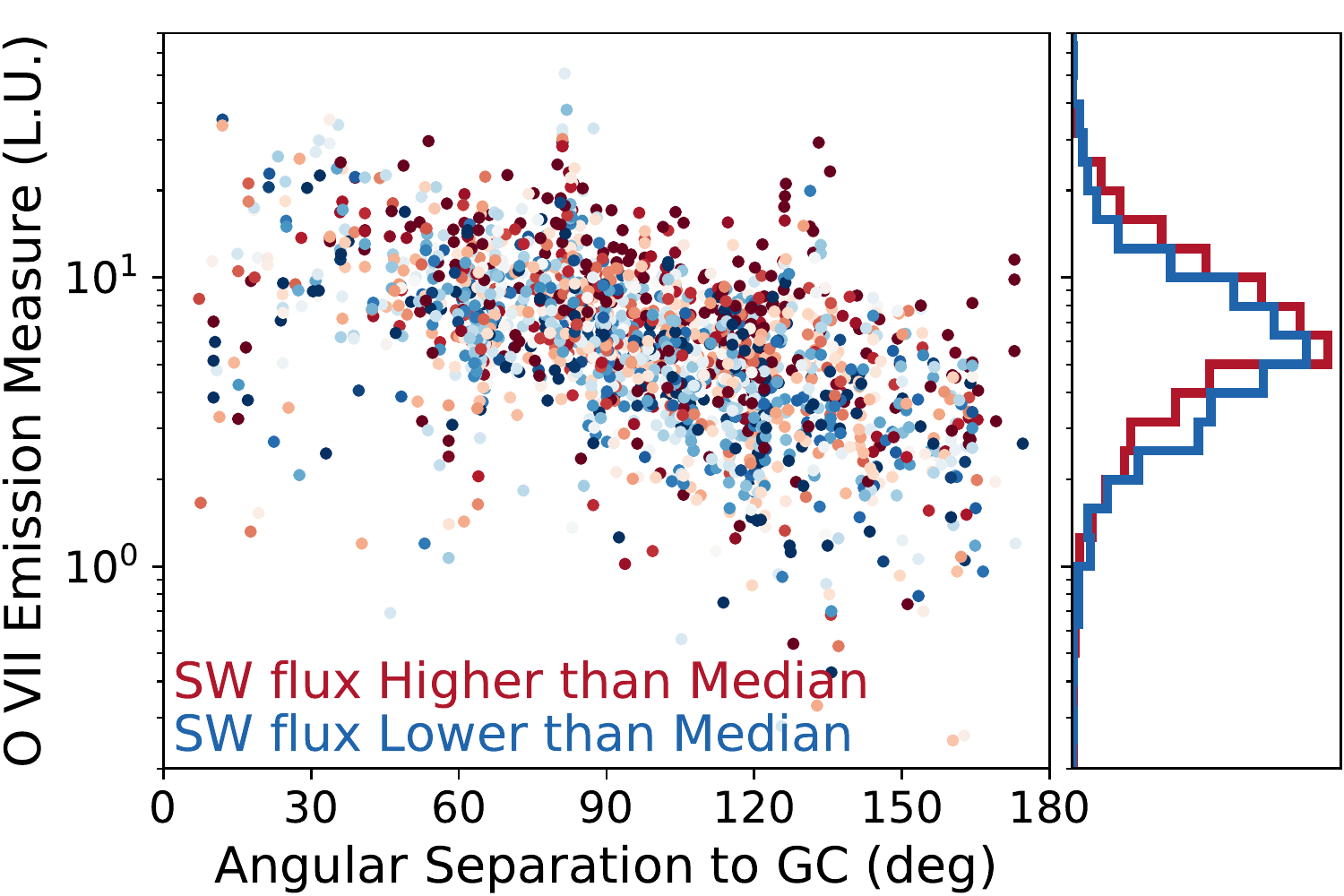}
\includegraphics[width=0.45\textwidth]{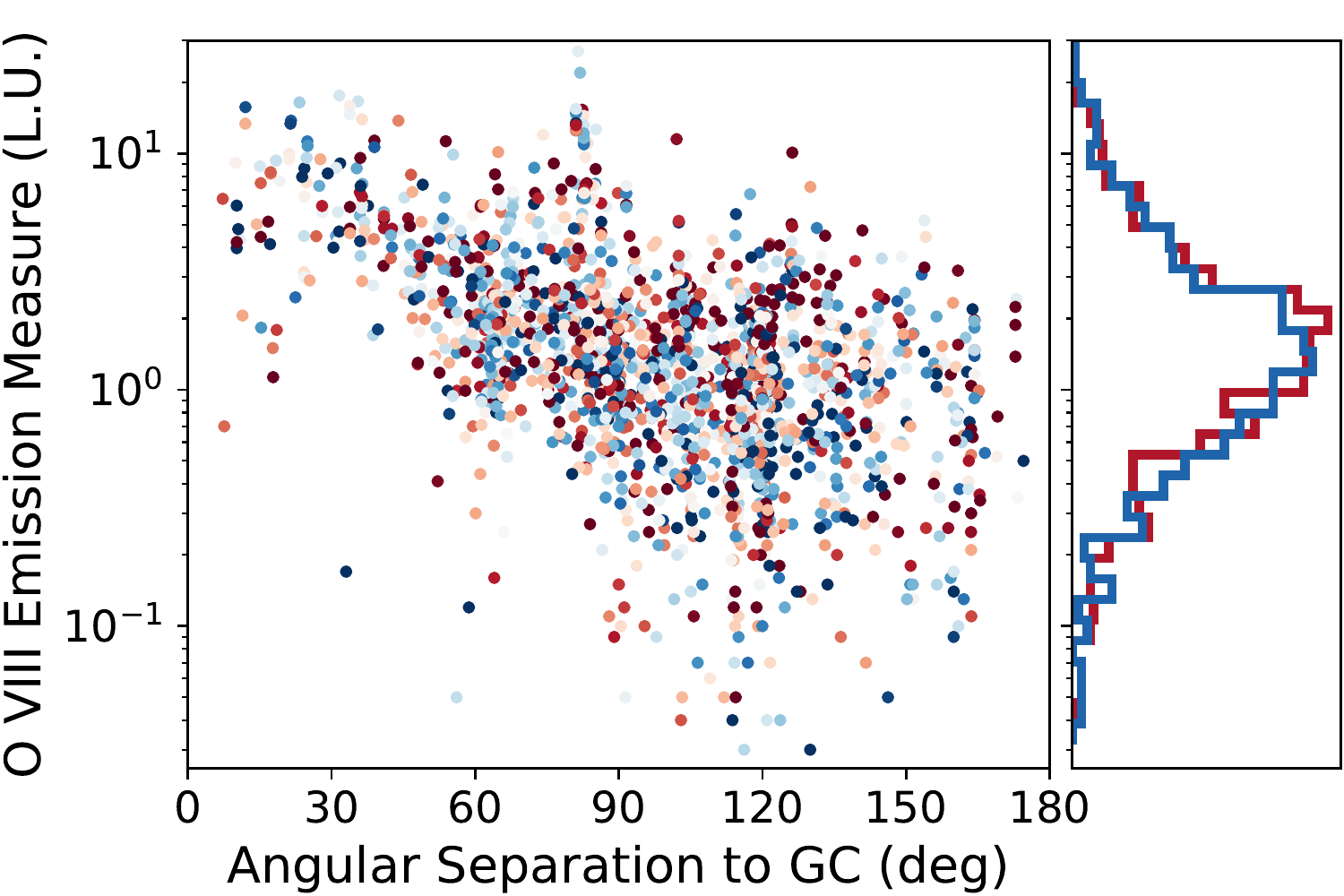}\\
\vspace{0.3cm}
\includegraphics[width=0.9\textwidth]{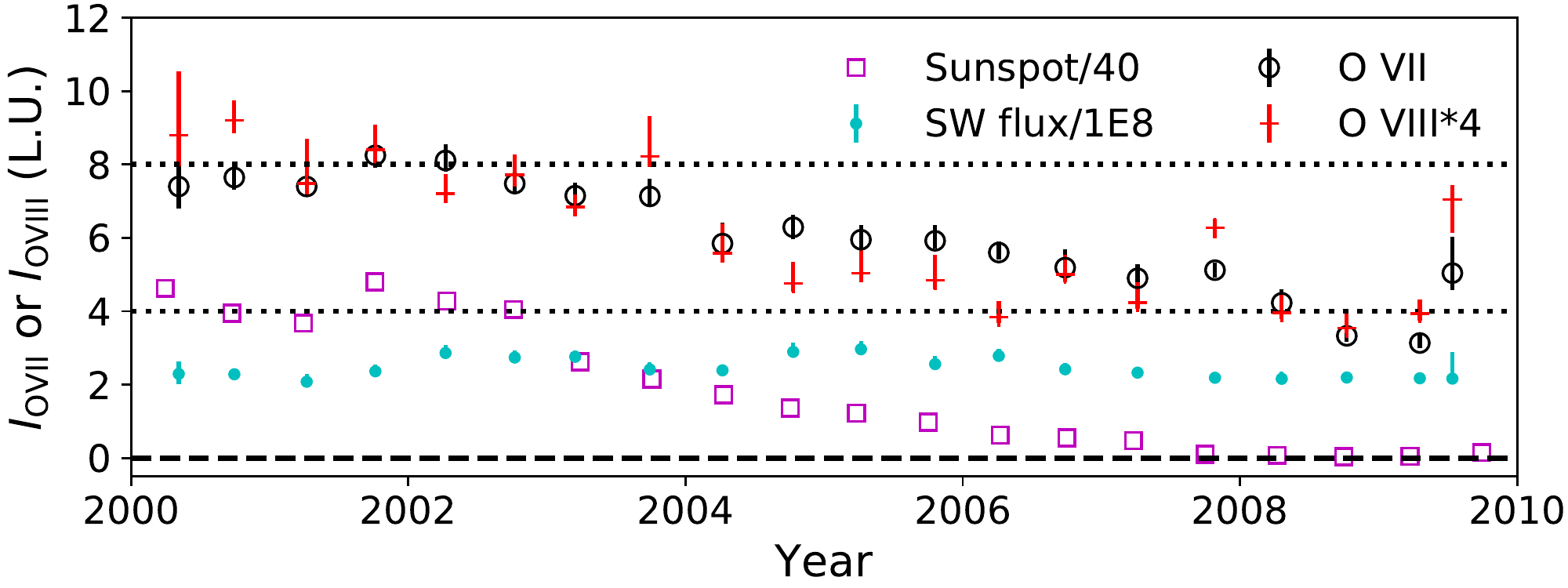}
\end{center}
\caption{{\it Top panels:} the \ion{O}{7} (left) and \ion{O}{8} (right) fluxes depend on the angular separation related to the Galactic center and the in-situ SW proton flux.
The points are color-encoded by the SW proton flux, and reddish points have higher SW proton fluxes.
The half-sample with higher SW fluxes has higher \ion{O}{7} ($3.7 \sigma$) and \ion{O}{8} fluxes ($0.8 \sigma$).
{\it Bottom panel:} the light curves of observed \ion{O}{7}, \ion{O}{8} (scaled up by 4), the in-situ SW proton flux (scaled down by $10^8$; calculated for individual observations by \citetalias{Henley:2012aa}), and the sunspot number (SSN; scaled down by 40).
Both \ion{O}{7} and \ion{O}{8} fluxes show significant declines corresponding to the SSN tracing the solar cycle in six-month long bins.
}
\label{obs}
\end{figure*}

\section{Data and Observed Correlations}

In this study we adopted the data from \citetalias{Henley:2012aa}, which reduced all archived {\it XMM-Newton} observations prior to 2010 August 4.
\citetalias{Henley:2012aa} measured the \ion{O}{7} and \ion{O}{8} line fluxes by disabling their emissions in {\tt APEC} or {\tt MEKAL} models \citep{Mewe:1985aa, Kaastra:1993aa, Smith:2001aa}, and adding two Gaussian emission lines to extract net line fluxes.
They adopted three steps to reduce contamination after masking out solar flares. 
First, they masked out point sources with $0.5 - 2.0$ keV flux $F_{\rm X}^{0.5-2.0 \rm~ eV} \geq 5\times10^{-14}\rm~erg~cm^{-2}~s^{-1}$ in the Second XMM-Newton Serendipitous Source Catalog \citep{Watson:2009aa} and extended sources by visual inspection.
Second, they filtered out observations with high extragalactic power-law fluxes at $2-5$ keV to exclude strong contamination from remaining point sources in the field, which resulted in a sample of 1868 sight lines.
Third, the local SWCX contamination is reduced by filtering out periods with high local SW proton fluxes $> 2 \times 10^8$ counts $\rm~ cm^{-2}$ (around the Earth obtained from the OMNIWeb database; \citealt{King:2005aa}), which reduces the sample size to 1003.

However, these steps do not ensure that the SWCX is entirely removed in the \citetalias{Henley:2012aa} low SW flux sample.
In Fig. \ref{obs}, we divide the full sample of 1868 sight lines into two subgroups based on the median of the SW flux.
The high SW flux sample leads to higher \ion{O}{7} fluxes with a median of $6.42\pm0.16$, while the low SW flux sample has a median of $5.57\pm 0.16$, which is a $3.7\sigma$ difference.
The $p$-value is $1.1\times 10^{-7}$ in the Kolmogorov-Smirnov (KS) test.
For \ion{O}{8}, the median values of $1.35\pm0.07$ and $1.25\pm0.09$ for high and low SW sample, respectively.
This is a $0.8 \sigma$ difference and the KS test $p$-value is 0.29.
This dependence of \ion{O}{7} and \ion{O}{8} fluxes on the local SW flux is evidence for the magnetospheric SWCX \citep{Cravens:2001aa}.

The heliospheric SWCX also introduces significant features in the \citetalias{Henley:2012aa} sample.
We resample the \ion{O}{7} and \ion{O}{8} line fluxes based on the observation date in bins of six months in Fig. \ref{obs}.
There is a significant decline of the \ion{O}{7} and \ion{O}{8} line fluxes from 2000 to 2010.
A KS test shows a $p$-value of $3.8\times10^{-15}$ on the two subsamples divided by observation date of 2005 for both \ion{O}{7} and \ion{O}{8}, which suggests a significance $>10 \sigma$.
The peak of the \ion{O}{7} flux is about 8 L.U. in 2002, and the minimum is about 4 L.U. around 2009.
This variation matches with solar cycle represented by the SSN, which has a maximum around 2000-2003, and a minimum around 2008.
Therefore, we suggest that this solar-cycle variation of the \ion{O}{7} and \ion{O}{8} fluxes is evidence for the heliospheric SWCX.
We also note that this long-term heliospheric SWCX variation over years is different from the previously-known LTE (i.e., the magnetospheric SWCX), because the average SW flux at the Earth does not vary significantly from 2000 to 2010 (Fig. \ref{obs}).

\section{A SWCX Empirical Model}

We constructed an empirical model of the overall SWCX variation in order to remove the bulk of the SWCX emission from the {\it XMM-Newton} observations, and to produce cleaner measures of the emission due to the MW hot halo.
We model the variation in the two types of SWCX using their expected behavior.
The magnetospheric SWCX is expected to be proportional to the SW flux, while the heliospheric SWCX is expected to be a function of solar activity.

The \ion{O}{7} and \ion{O}{8} SWCX lines are produced by charge exchange with O$^{+7}$ and O$^{+8}$ in the SW.
The density of those two ions within the line of sight depends upon both the SW flux (typically given as the proton flux) and the relative abundances of the ions with respect to the protons. 
Those abundances are a function of solar activity (e.g., the slow solar equatorial flow or the fast solar polar flow).
The slow flow has higher O$^{+7}$ and O$^{+8}$ abundances than the fast flow \citep{Kuntz:2019aa}.
During the solar minimum, the two flows are distinctly segregated, while at the solar maximum, the flows are intermixed.
Thus, the overall properties of the O$^{+7}$ and O$^{+8}$ abundance will vary with the solar cycle.
While some measures of SW O$^{+7}$ and O$^{+8}$ fluxes have been available from ACE, they do not have high signal-to-noise ratios, and those measures are representative of the solar equatorial flow around the Earth, which will be different from the solar polar flow.
Therefore, we do not use the O$^{+7}$ and O$^{+8}$ fluxes in the empirical model.

The magnetospheric SWCX (e.g., the {\it ROSAT} LTE strength) is closely correlated with the SW flux around the Earth \citep{Cravens:2001aa, Kuntz:2015aa}. 
Fig. \ref{obs} reiterates this dependence, showing that the \ion{O}{7} and \ion{O}{8} line strengths are greater for observations during high SW flux periods than for observations during low SW flux periods.
Therefore, in the empirical model, we assume that the magnetospheric SWCX is proportional to the local SW flux.

The heliospheric SWCX depends upon the observing line of sight (through the equatorial flow or the polar flow).
These solar flows have dependences on the solar cycle (e.g., weaker polar flows at the solar minimum; \citealt{McComas:2008aa}).
The standard measure of solar activity within the solar cycle is the SSN, which is tightly correlated with a number of other properties of the solar cycle, such as the magnetic field \citep{Smith:2014ab}. 
Therefore, we assume that the heliospheric SWCX is proportional to the SSN.
We note that, as the SW flux is weakly correlated with the SSN, our measures of the magnetospheric and heliospheric SWCX are not entirely decoupled.
However, this effect will not be noticeable in the following analysis.

For the heliospheric SWCX, we first smooth the SSN curve with a Gaussian function of $\sigma=1$ yr to reduce the uncertainty introduced by the scattering and inter-cycle periodicity of the SSN (periods up to half a year; \citealt{Joshi:2006aa}).
The inclusion of the 1-year smoothing factor will not change the long-term SSN variation, because the SSN peak can be approximated by a Gaussian with a width of $\approx 3$ year.
The additional 1-year smoothing only increases the width by $<5\%$, which cannot be distinguished by the current data.
Our formulation also allows a lag ($\tau$) between the SSN and observed \ion{O}{7} and \ion{O}{8} SWCX.
This lag has three possible physical origins discussed in Section \ref{section:lag}.

The contribution by the MW to the \ion{O}{7} and \ion{O}{8} fluxes is non-negligible.
The MW contribution depends on the Galactic coordinates, showing an dependence on the angular separation to the Galactic center (GC; Fig. \ref{obs}), but does not have any dependence on the solar cycle or the local SW flux.
Therefore, there is no correlation expected between the MW contribution and the magnetospheric or heliospheric SWCX.
We assume a constant value for the Galactic contribution to the \ion{O}{7} and \ion{O}{8} emission for this temporal analysis.
The spatial distribution of MW hot gas emission will be discussed in Section \ref{section:MW}.
In this study we mask out sight lines with \ion{H}{1} column densities $>2\times10^{21}\rm~cm^{-2}$ (adopted from the HI4PI survey; \citealt{HI4PI-Collaboration:2016aa}) in the final sample with 1602 observations.
These sight lines are at low Galactic latitudes, often crossing bright structures, and contain emission that is different than the hot halo emission at higher Galactic latitudes.

Thus, the empirical model has three components.
The observed \ion{O}{7} and \ion{O}{8} are calculated as
\begin{equation}
I_{\rm m} =  A_{\rm mag}F_{\rm SW} + B_{\rm helio} f(N_{\rm SSN}, \sigma, \tau)  + I_{\rm MW},
\end{equation}
where $F_{\rm SW}$ is the local SW flux (adopted from \citetalias{Henley:2012aa}) for individual observations, and $f(N_{\rm SSN}, \sigma, \tau)$ is the smoothed ($\sigma$) and lagged ($\tau$) SSN.
The normalization factors ($A_{\rm mag}$, $B_{\rm helio}$, and $I_{\rm MW}$ for magnetospheric, heliospheric, and MW, respectively) are different for \ion{O}{7} and \ion{O}{8}, but we assume that \ion{O}{7} and \ion{O}{8} share the same lag time ($\tau$).
Therefore, in the empirical model, there are seven free parameters and one fixed parameter (smoothing width $\sigma = 1$ yr).

\begin{figure*}
\begin{center}
\includegraphics[width=0.7\textwidth]{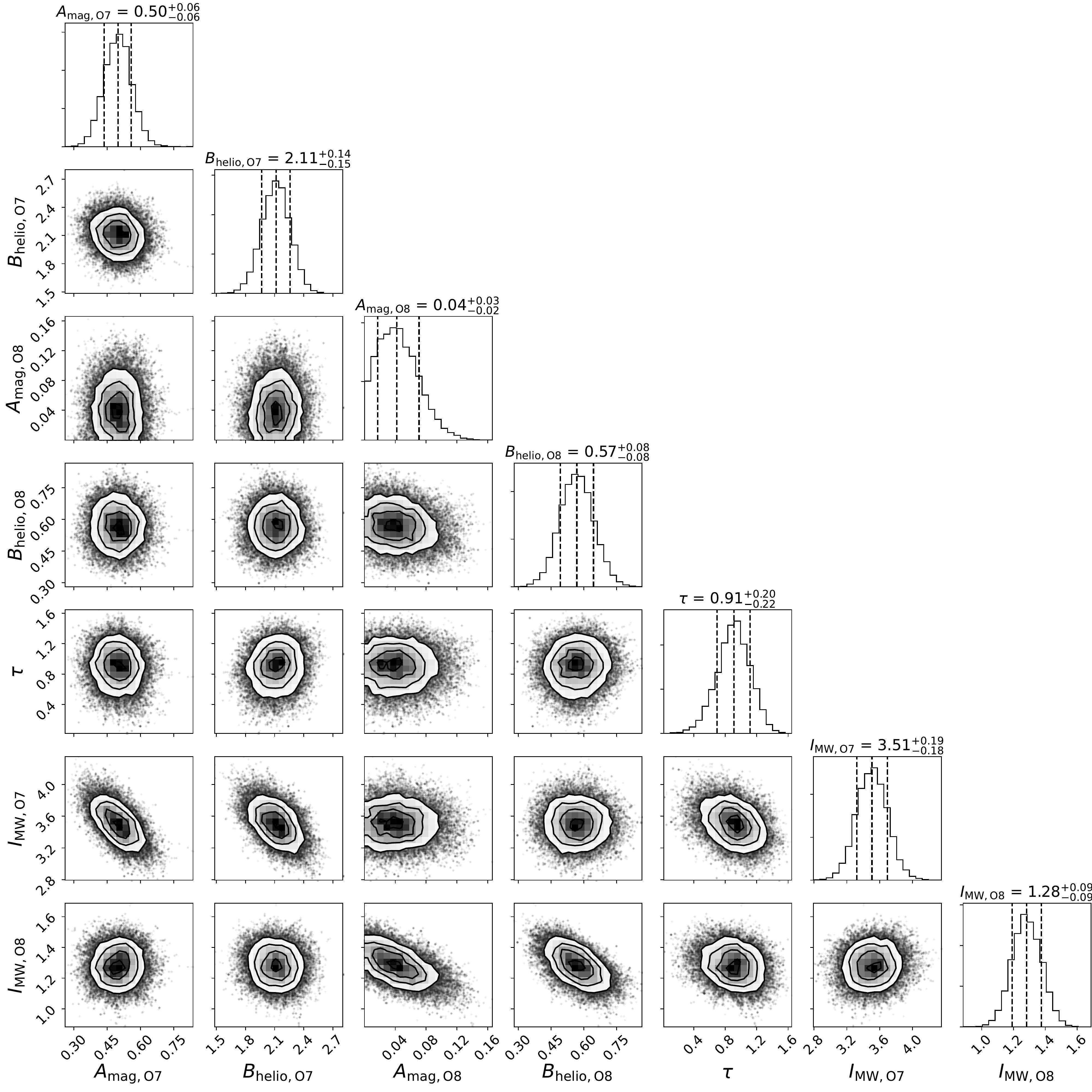}
\end{center}
\caption{Best parameters of the empirical SWCX model extracted from {\it emcee}. $A_{\rm mag}$ and $B_{\rm helio}$ are the normalization factors for the magnetospheric (depending on the in-situ SW proton flux) and heliospheric (depending on the SSN) SWCX, which are different for \ion{O}{7} and \ion{O}{8}. The lag time between the observed line measurements and the SSN is assumed to be the same for both \ion{O}{7} and \ion{O}{8}.}
\label{params_corner}
\end{figure*}

We adopt the Bayesian optimization to obtain the best parameters in the model.
Specifically, the likelihood is calculated using the residuals in the logarithmic scale, because both the MW residual and the SWCX scatter are expected to be lognormal:
\begin{equation}
\ln p = \sum \frac{(\log I - \log I_{\rm m})^2}{\sigma_m^2+\sigma_p^2} - \frac{1}{2} \ln (\sigma_m^2+\sigma_p^2), 
\end{equation}
where $I$ is the observed \ion{O}{7} or \ion{O}{8} line emission measurement, and $I_{\rm m}$ is the model defined in Equation (1).
The observational uncertainty ($\sigma_m^2$) is adopted from \citetalias{Henley:2012aa}, which is calculated by combining the systematic uncertainty and the measurement uncertainty.
The parameter $\sigma_p^2$ is an empirical patchiness parameter to account for the scatter in residuals (mainly due to the MW spatial variation), which is different for \ion{O}{7} and \ion{O}{8}.

The best parameters are obtained with {\it emcee} (\citealt{Foreman-Mackey:2013aa}; shown in Fig. \ref{params_corner}).
The fitted scaling relations for the magnetospheric and heliospheric SWCX \ion{O}{7} or \ion{O}{8} are 
\begin{eqnarray}
I_{\rm O7, mag} & = & (0.50\pm 0.06) F_{\rm SW, 8}, \notag \\
I_{\rm O8, mag} & = & (0.04_{-0.02}^{+0.03}) F_{\rm SW, 8}, \notag \\
I_{\rm O7, helio} & = & (2.11_{-0.15}^{+0.14}) f(N_{\rm SSN}, \sigma, \tau), \notag\\
I_{\rm O8, helio} & = & (0.57\pm 0.08) f(N_{\rm SSN}, \sigma, \tau), \notag\\
I_{\rm O7, MW} & = & 3.51_{-0.18}^{+0.19}, \notag\\
I_{\rm O8, MW} & = & 1.28\pm 0.09,
\end{eqnarray}
where $F_{\rm SW, 8}$ is the SW proton flux in units of $1\times10^8$ count s$^{-1}$.
The smoothing timescale is fixed to one year, and the lag is $\tau = 0.91_{-0.22}^{+0.20}$ yr ($11_{-3}^{+2}$ months), which differs from zero at $4 \sigma$ significance.

\begin{figure*}
\begin{center}
\includegraphics[width=0.8\textwidth]{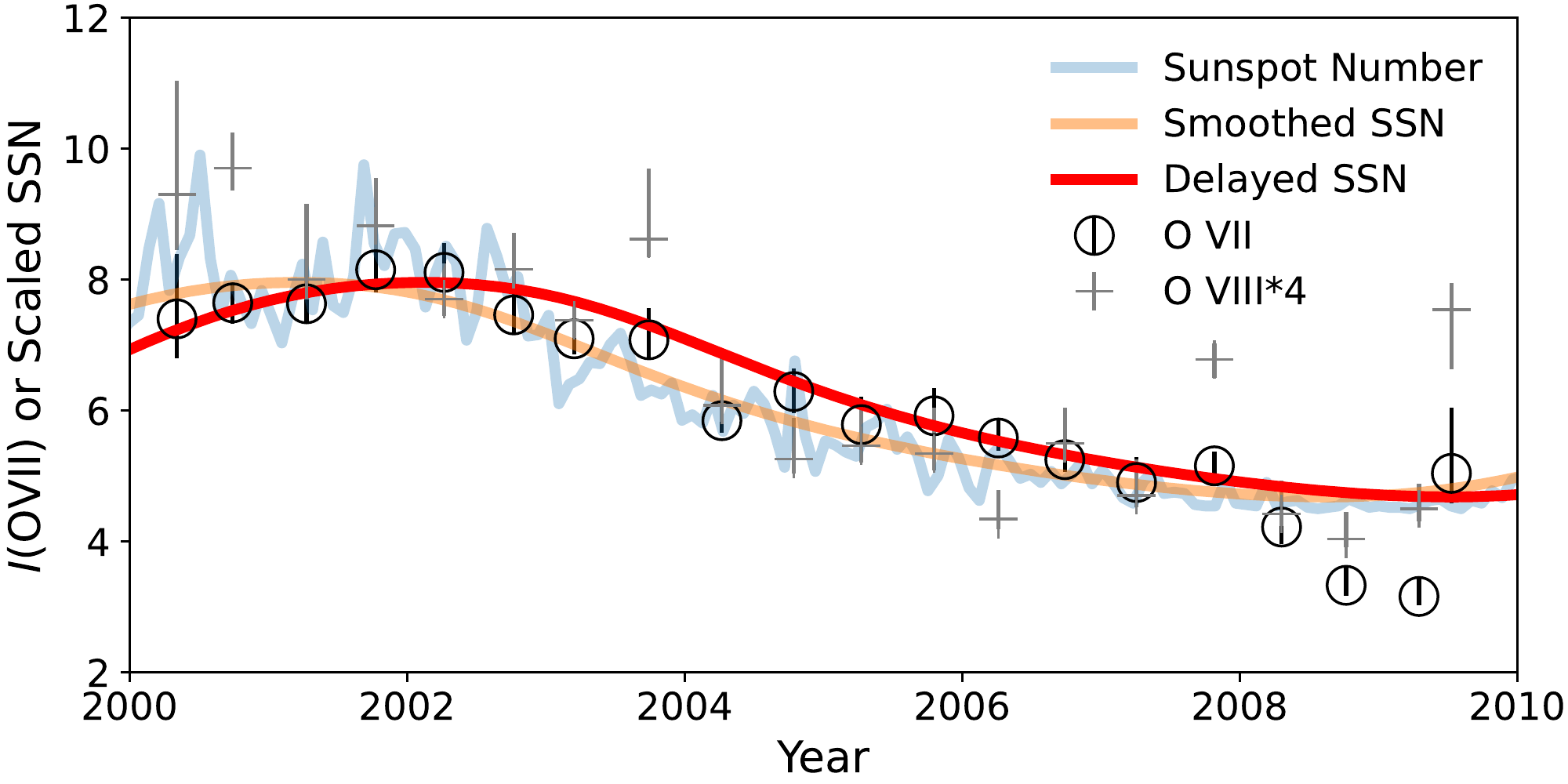}
\end{center}
\caption{A comparison between the smoothed, lagged SSN and the observed \ion{O}{7} and \ion{O}{8} fluxes. The red line lags by 0.91 yr compared to the orange line, which is the SSN light curve smoothed with a Gaussian function with $\sigma =1$ yr.
}
\label{delay}
\end{figure*}

\section{Implication and Discussion}
\subsection{The Lag between the SSN and the SWCX}
\label{section:lag}
In Fig. \ref{delay}, we scale the SSN to match with the observed \ion{O}{7} flux and show this lag.
This lag is dominated by \ion{O}{7}, while \ion{O}{8} is too uncertain to determine the lag itself.

There are three possible origins for this lag.
First, the heliospheric SWCX is expected to have a lag due to the transit time for the SW to reach the heliosphere.
The radius of the heliosphere is about 120 AU \citep{Richardson:2019aa}, so the transit time of the SW is about 1.4 years to reach the heliopause, considering the typical SW velocity of $400 \kms$
More accurate estimation of the transit time for the SWCX involves the neutral gas distribution, which suggests that $>80\%$ of SWCX emission are within 20 AU \citep{Kuntz:2019aa}, which leads a SW transit time of $0.2-0.3$ yr.

Second, recent studies have revealed that the neutral gas distribution is non-stationary showing corresponding variation with the SW propagation \citep{Koutroumpa:2021aa}.
Increased SW flux could sweep up or ionize the neutral gas close to the Sun.
The variation of the neutral gas density distribution leads to a temporal variation of the integration of the neutral gas, which lags behind the SW variation, and further increases the lag between the SSN and the SWCX.

The third possibility is that there is an intrinsic lag between the SSN and the high ionization state ion flux in the SW corresponding to the solar cycle.
Previous studies confirmed $\approx 1 -15$ month lags between different tracers of the solar cycle (e.g., \citealt{Bachmann:1994aa, Temmer:2003aa, Ramesh:2014aa}).
It is possible that the high-energy solar activities generating $\rm O^{7+}$ and $\rm O^{8+}$ intrinsically lag the SSN, which accounts for an extra lag other than the transit time.

\subsection{Earth Position-related Variation}
\label{section:yearly}
There is an Earth position-related (seasonal) variation of the \ion{O}{7} sample in the {\it XMM-Newton} archive, while the \ion{O}{8} sample does not show this variation due to its lower signal-to-noise ratio.
This seasonal variation is regulated by two factors.
First, the neutral IPM is anisotropic due to the solar motion relative to the nearby ISM \citep{Baranov:1993aa, Opher:2020aa}.
The up-wind direction is roughly near the GC direction ($l,~b) = (3^\circ,~ 16^\circ$). 
In this direction, the neutral gas forms a high hydrogen density region \citep{Lallement:1984aa}.
In the down-wind direction, although the hydrogen density is relatively lower, the helium density is enhanced due to gravitational focusing of helium by the Sun, which is effective because of the higher mass of helium \citep{Dalaudier:1984aa}.
These over-dense regions lead to higher SWCX emissivities when the Earth passes through them.
Thus, the heliospheric SWCX will have a seasonal variation that is closely aligned with the Galactic center/anti-center axis.
Second,  {\it XMM-Newton} mostly observes targets within $\pm10^\circ$ perpendicular to the Sun-Earth direction (selection of ecliptic longitudes), which is required to ensure sufficient energy supply and thermal stability of the spacecraft.
Therefore, the observation pointing directions also have seasonal variation.

\begin{figure*}
\begin{center}
\includegraphics[width=0.9\textwidth]{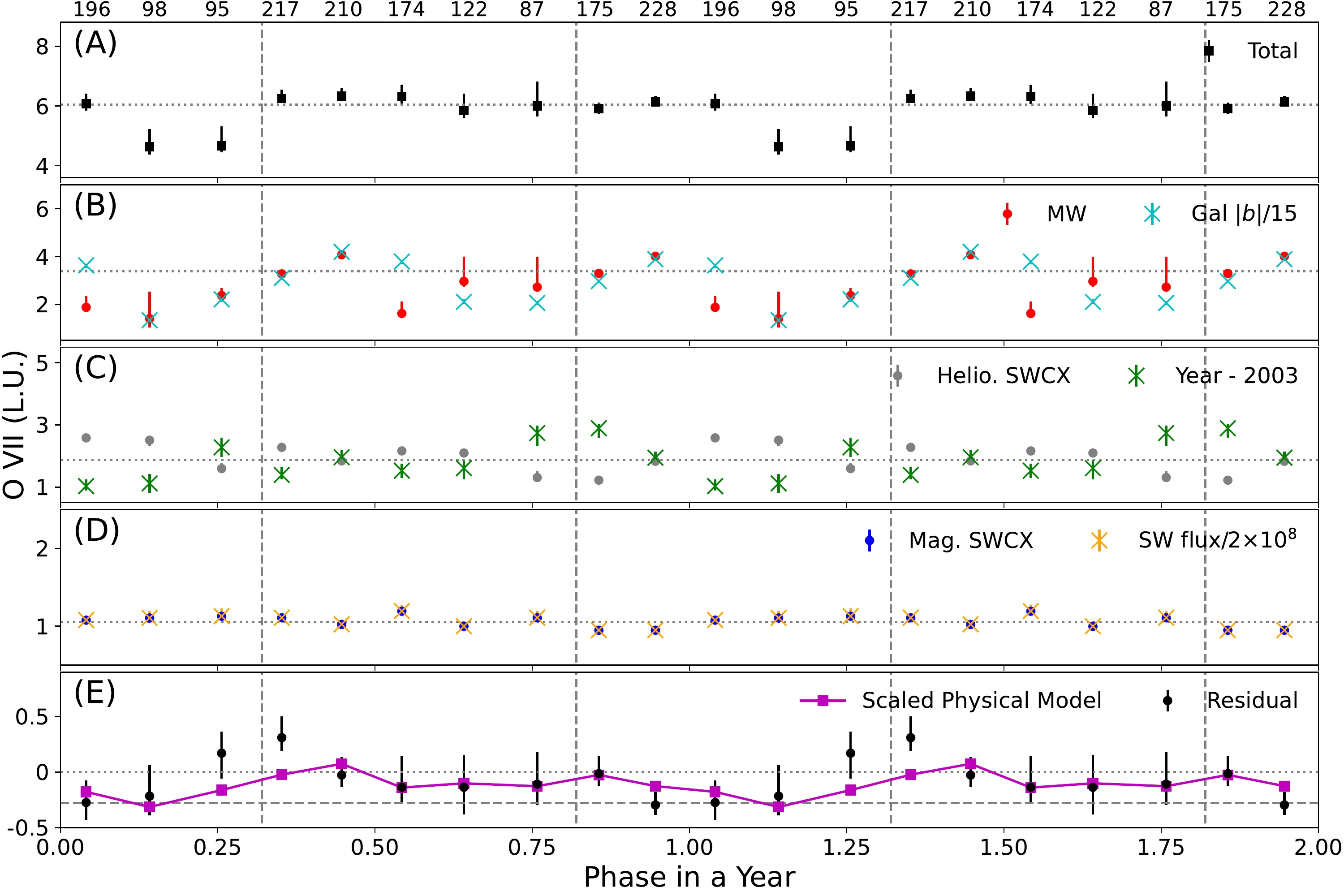}
\end{center}
\caption{The seasonal variation of different components and residuals (subtracting SWCX and MW contributions) in the empirical model. In this plot, the phase of $1-2$ repeats the phase $0-1$ for illustration. The values on the top axis are the number of observations in each bin. 
{\bf Panel (A)}: the observed \ion{O}{7} flux. 
{\bf Panel (B)}: the seasonal variation of the MW \ion{O}{7} model (red dots) is dominated by the {\it XMM-Newton} observing strategy, which leads to a selection bias of Galactic latitude (cyan crosses; scaled down by a factor of 15).
{\bf Panel (C)}: the heliospheric SWCX (gray dots) shows a seasonal variation due to the observation date (green crosses; see Section \ref{section:yearly}), because a later observation date leads to a lower SSN and modeled heliospheric SWCX.
A value of 2003 is subtracted from the average observation date to compare with the heliospheric SWCX flux.
{\bf Panel (D)}: the magnetospheric SWCX (blue dots) is assumed to be proportional to the local SW flux (orange crosses), which does not show a seasonal variation.
{\bf Panel (E)}: the residuals show a $3\sigma$ seasonal variation after subtracting the two SWCX and the MW models. This seasonal variation has two peaks (dashed vertical lines) close to the up-wind and down-wind (helium focusing cone) directions. These two peaks are dominated by the heliospheric SWCX variation and modulated by the {\it XMM-Newton} observing strategy. We compared the observed residuals with the model prediction introduced in the Appendix (the magenta line in the lowest panel).
The gray dashed line at $-0.3$ L.U. shows the minimum for the seasonal variation of the heliospheric SWCX, which should be corrected for the MW \ion{O}{7} fluxes (see Section \ref{section:MW}).
}
\label{yearly_phase}
\end{figure*}

This seasonal variation is not considered in the empirical model, and, indeed, is mostly removed by our six month binning.
Instead, we examine and correct this seasonal variation using the residual of the empirical model after subtracting long-term temporal variation of the two modeled SWCX components and the MW contribution.
The two SWCX components are calculated from the empirical model.
The MW contribution is not the constant value in the empirical model, instead, we calculate a position-dependent MW contribution.

We consider the spatial distribution of the MW hot halo emission, after subtracting the SWCX empirical models and correcting for the absorption ({\tt phabs} absorption in {\tt XSPEC}).
The MW \ion{O}{7} and \ion{O}{8} line fluxes have a strong dependence on the angular distance to the GC \citep[e.g., ][]{Miller:2015aa}.
Therefore, the MW contribution is calculated as the interpolation of the angular radial profile related to the GC.
For this empirical MW model, we consider the north-south asymmetry, where the northern sky is brighter than the southern sky by about 1 L.U. \citep{Qu:2021aa}.
Here, the empirical MW model still has contributions from the seasonal variation of the heliospheric SWCX, because the seasonal variation is not included in the empirical SWCX model.
This seasonal variation has a non-zero mean, so this MW model is overestimated, which will be corrected in Section \ref{section:MW}.

In Fig. \ref{yearly_phase}, we show the seasonal variation of the three empirical model components.
The MW and the heliospheric SWCX have significant seasonal variations ($>3\sigma $), while the magnetospheric SWCX is a constant within uncertainty.
First, the seasonal variation of the MW contribution is dominated by the {\it XMM-Newton} observing constraints, which samples different distances from the GC at different times of year (panel B in Fig. \ref{yearly_phase}).
Second, for the heliospheric SWCX model, we do not expect temporal variations shorter than one year (including the seasonal variation), because the SSN is smoothed with one year in the empirical model.
However, we find that the average observation dates in different seasonal bins are not constant (Fig. \ref{yearly_phase}), which leads to the observed seasonal variation for the heliospheric SWCX.
From 2000 to 2010, the solar activity evolved from its maximum to its minimum, so a later average observation date leads to a lower heliospheric SWCX (panel C in Fig. \ref{yearly_phase}).
Third, the magnetospheric SWCX does not show a seasonal variation, as expected (panel D in Fig. \ref{yearly_phase}).
The X-ray and soft proton flare filter in \citetalias{Henley:2012aa} data reduction also filter out high local SW flux periods, which leads to the flat magnetospheric SWCX.

\begin{figure*}
\begin{center}
\includegraphics[width=0.45\textwidth]{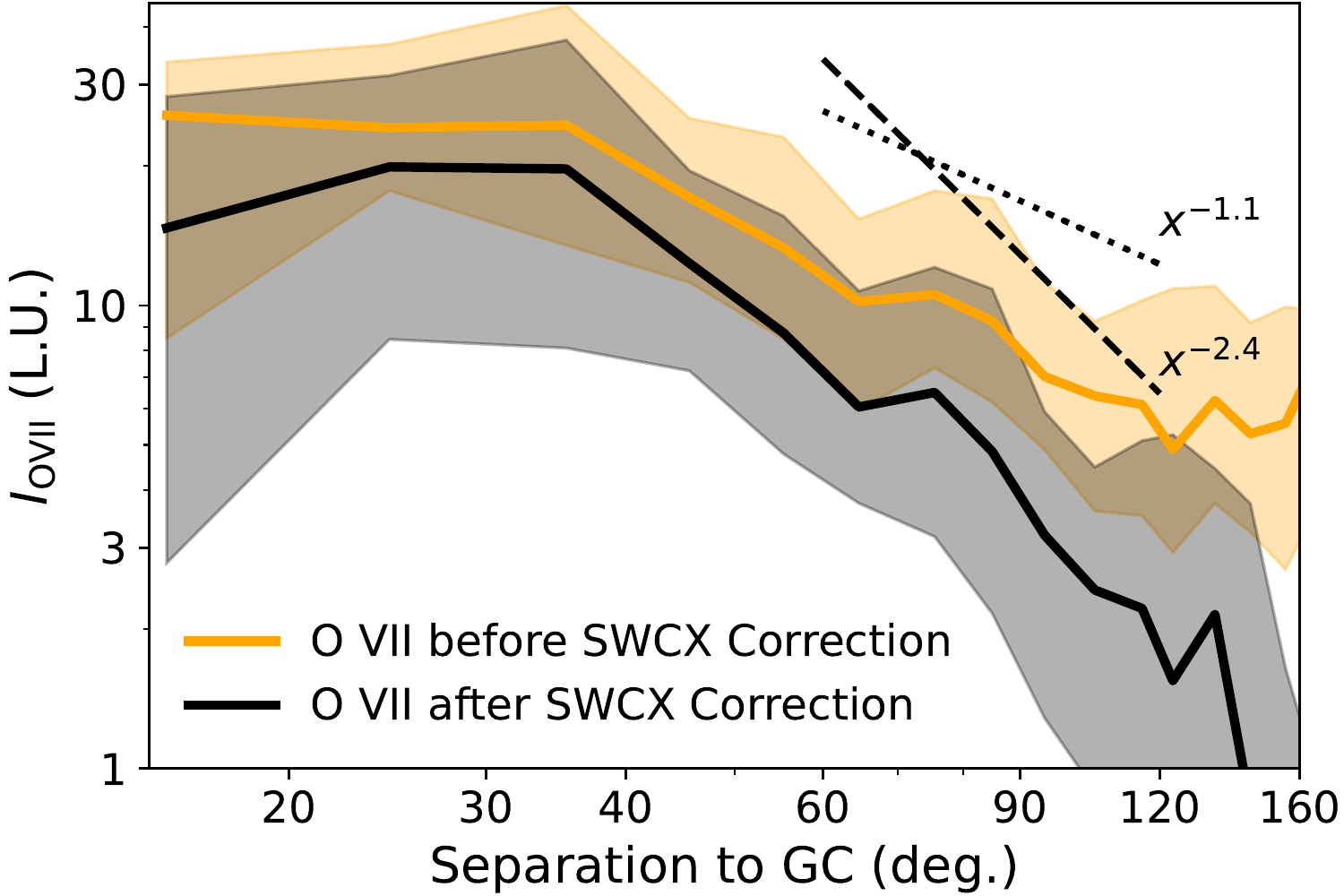}
\includegraphics[width=0.45\textwidth]{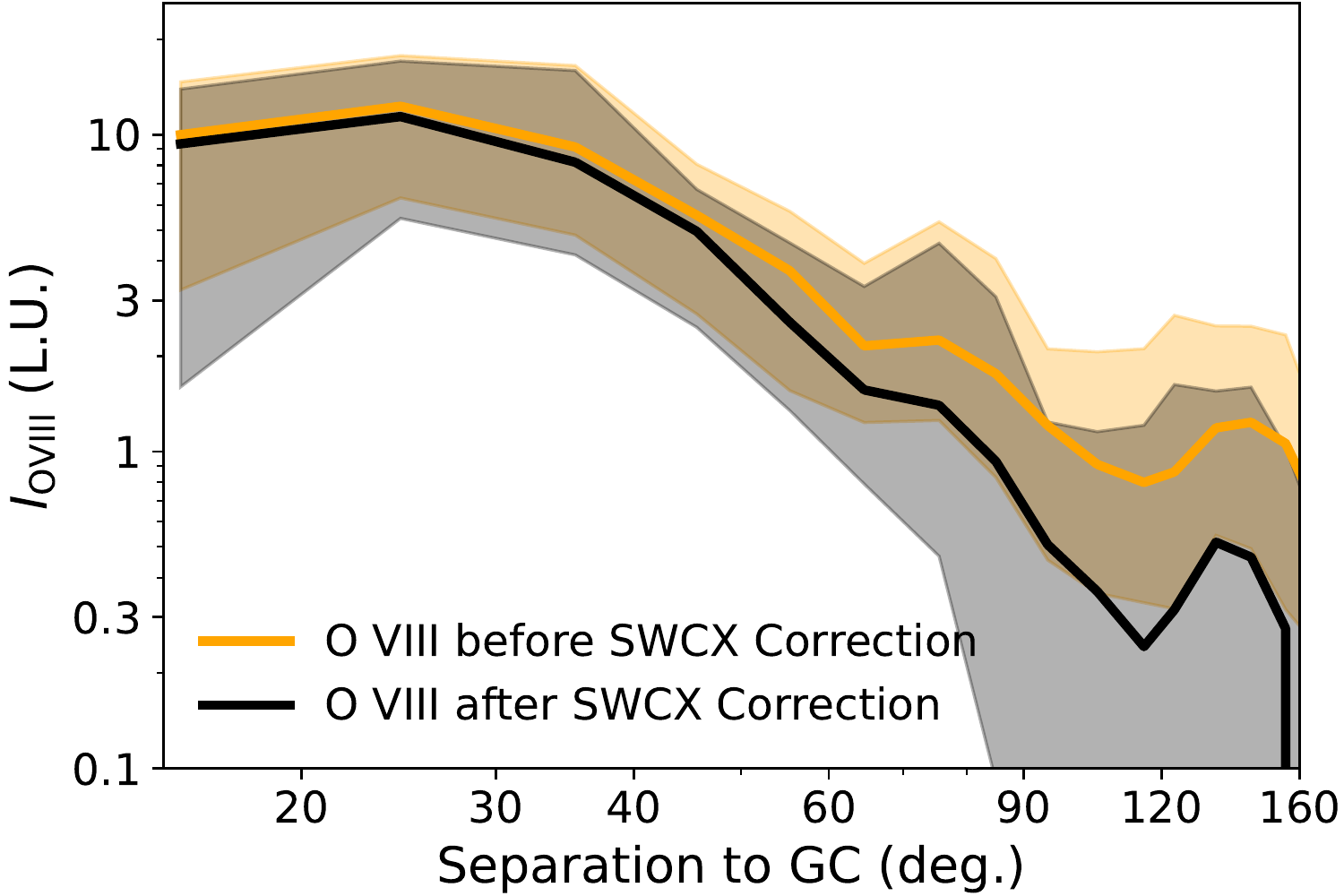}\\
\vspace{0.3cm}
\includegraphics[width=0.9\textwidth]{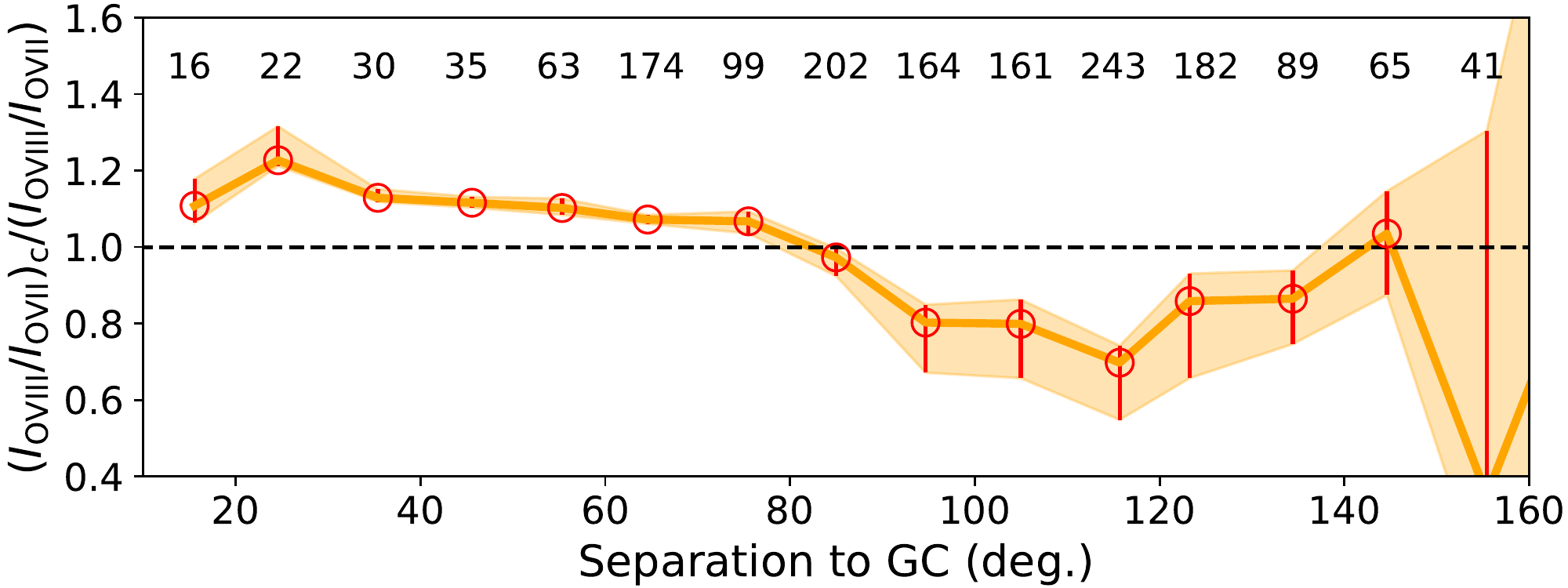}
\end{center}
\caption{{\it Top panels:} the SWCX correction for MW \ion{O}{7} (left) and \ion{O}{8} (right). The average line fluxes are reduced by $50\%$ and $30\%$ for \ion{O}{7} and \ion{O}{8}, respectively. The correction is more influential in the outskirts, and steepens the decreasing slope at large radii. Two power laws are shown to approximate the decrease of  Galactic \ion{O}{7} emission at larger radii. 
{\it Bottom panel:} The ratio of the corrected to uncorrected temperature-sensitive line ratio $I_{\mbox{\ion{O}{8}}} / I_{\mbox{\ion{O}{7}}}$, where larger ratios ($>1$) imply higher temperatures after the SWCX correction.
The number of observations is labeled above each bin.
}
\label{mw}
\end{figure*}

The final residual of \ion{O}{7} [$I_{\rm total} - (I_{\rm mag} + I_{\rm helio} + I_{\rm MW})$; panel A-(B+C+D) in Fig. \ref{yearly_phase}] is shown in the last panel in Fig. \ref{yearly_phase}, which shows a tentative ($\approx 3\sigma$) seasonal variation.
This seasonal variation has two peaks at seasonal phase of 0.35 and 0.85, which are close to the up-wind (0.42) and down-wind directions (0.92).
The up-wind peak is slightly stronger than the downwind peak.
We adopt a normalized simulation model (see the Appendix) to examine whether this observed seasonal variation in the residuals is the seasonal variation of the heliospheric SW.
This simulation model only accounts for the neutral gas distribution, while the variation in the $\rm O^{7+}$ SW flux is not included in this model.
This normalized model is scaled to match with the observed residual with an offset of $1.2\pm 0.7$ L.U., which is the average heliospheric SWCX (within $1\sigma$ of the empirical model of $1.8\pm 0.3$ L.U.).
We suggest that the observed residual is the seasonal variation of the heliospheric SWCX.

This seasonal variation residual varies from $-0.3$ to $+0.3$ L.U. (Fig. \ref{yearly_phase}). 
However, this residual seasonal variation should have a non-negative minimum instead of $-0.3$ L.U. in Fig. \ref{yearly_phase}, because the empirical heliospheric SWCX model does not include this seasonal variation.
Therefore, an average flux of $\approx 0.3$ L.U. has been attributed to the current MW empirical model, which will be corrected in Section \ref{section:MW}.

\subsection{The SWCX-Clean MW Hot Gas Emission}
\label{section:MW}

The MW hot halo emission extracted in Section \ref{section:yearly} is still contaminated by the heliospheric SWCX seasonal variation.
To correct this contamination, we subtract another 0.3 L.U. together with observed residual seasonal variations from the MW emission.

In Fig. \ref{mw}, we show the MW hot gas emission before and after the SWCX correction.
The average MW fluxes are reduced by $ 50\%$ for \ion{O}{7}  (the mean of 10.0 L.U. to 5.4 L.U.) and and $30\%$ for \ion{O}{8} (2.5 L.U. to 1.7 L.U.), compared to previous studies where the SWCX component is not accounted for (e.g., \citealt{Gupta:2012aa, Miller:2015aa}).
An order-of-magnitude estimation suggests that the MW hot gas mass and the hot gas mass is overestimated by $20-30\%$.
Furthermore, the SWCX-correction is more important in the outskirts, where the MW emission is low.
Previously, the MW hot gas distribution (within 50 kpc) is determined to be a power-law with a slope of $1.2-1.5$ by assuming a constant temperature of $2\times 10^6$ K in \citep{Miller:2015aa, Li:2017aa}.
For both \ion{O}{7} and \ion{O}{8}, our new samples have steeper decline, which determines the hot gas density distribution, and will further reduce the mass that may be hosted at large radii.

The SWCX correction also changes the temperature determination.
The MW hot gas temperature is declining from the inner region dominated by the hot gas bubbles Fermi and eROSITA bubbles to the outskirts \citep{Su:2010aa, Miller:2016ab, Predehl:2020aa}.
The \ion{O}{8}/\ion{O}{7} ratio, as the tracer of the temperature, varies from $0.6-0.7$ (inner $20^\circ$) to about $0.1-0.2$ (outskirts).
In the empirical model, the SWCX \ion{O}{8}/\ion{O}{7} ratio is $0.20_{-0.06}^{+0.03}$.
Therefore, the temperature is underestimated in the inner region and overestimated in the outskirts if the SWCX contribution is not accounted for.
In the lower panel Fig. \ref{mw}, we show the ratio of the \ion{O}{8}/\ion{O}{7} ratio before and after the SWCX correction, which shows a transition point at $80^\circ$ for the temperature correction.
Another trend is that the ratio increases again beyond $120^\circ$.
This increase indicates that a higher temperature around the MW disk, as the angular separation increases, the selected region will be more and more concentrated to the anti-GC direction in the disk.

The SWCX correction of the \citetalias{Henley:2012aa} sample improves the determination of both density distribution and the temperature profile of the MW hot gas.
The detailed modeling of the MW hot gas will be presented in a future work.

~\\

The authors would like to thank the anonymous referee, Eric Qu\'eMerais, Jiang-Tao Li, and Susan Lepri for their valuable input and thoughtful discussions on this work.
Z. Q. acknowledges {\it Astropy} \citep{Astropy-Collaboration:2013aa} and {\it emcee} \citep{Foreman-Mackey:2013aa}.
Z. Q. and J.N. B. are supported by NASA ADAP grant AWD012791 and the University of Michigan.
D. K.'s heliospheric modeling work is supported by CNES, and performed with the High Performance Computer and Visualisation platform (HPCaVe) hosted by UPMC-Sorbonne Universit\'e.
P. K. is supported by NASA grant 80NSSC20K0398.

\appendix

The SWCX emissivity depends on the neutral gas distribution within the heliosphere and the strength of the high-energy SW (i.e., $\rm O^{7+}$ and $\rm O^{8+}$).
The SWCX reaction between SW ions $\rm X^{q+}$ and heliospheric neutrals N (H or He) is
\begin{center}
\ce{$\rm X^{q+}$ + N -> $\rm X^{(q-1)+}$ + $\rm N^{+}$ + $\gamma$},
\end{center}
In this work, we focus on oxygen ions (X $=$ O) for parent charge states $q = 7, 8$, and $\gamma$ refers to the resulting \ion{O}{7} triplet ($E\approx 0.57$ keV) and \ion{O}{8} Ly$\alpha$ ($E\approx0.65$ keV) lines, respectively.

The simulated total flux along the line-of-sight is calculated as:
\begin{equation}
I = \frac{1}{4\pi} \int_{s=0}^\infty F_{\rm X^{q+}}(R) \left[n_{\rm H}(\lambda, \beta, R) \sigma^{\rm X^{q+}}_{\rm H} Y^{\rm X^{q+}}_{\rm H} + n_{\rm He}(\lambda, \beta, R) \sigma^{\rm X^{q+}}_{\rm He} Y^{\rm X^{q+}}_{\rm He} \right] {\rm d} s
\end{equation}
where $F_{\rm X^{q+}}(R) = F_{\rm X^{q+}}({\rm 1~ AU})\times({\rm 1~ AU}/R)^2 $ is the ion flux depending on the radial distance to the Sun ($R$), and $n_{\rm N}(\lambda, \beta, R)$ is the neutral density for H and He in the interplanetary space, dependent on direction in ecliptic coordinates ($\lambda, \beta$), and  $R$.
The parameter $\sigma^{\rm X^{q+}}_{\rm N}$ is the SWCX cross-section for reactions between ion $\rm X^{q+}$ and neutral N, and  $Y^{\rm X^{q+}}_{\rm N}$ is the corresponding line emission probability following the specific SWCX reaction.

The high-energy ion flux $F_{\rm X^{q+}}({\rm 1~ AU})$ has large uncertainties because only the in-situ flux can be measured at $1 \rm~ AU$ by current instruments.
Therefore, we isolate the ion flux from the SWCX calculation in the Equation (2), and define the normalized model:
\begin{equation}
Q = \frac{1}{4\pi} \int_{s=0}^\infty \frac{1}{R^2} \left[n_{\rm H}(\lambda, \beta, R) \sigma^{\rm X^{q+}}_{\rm H} Y^{\rm X^{q+}}_{\rm H} + n_{\rm He}(\lambda, \beta, R) \sigma^{\rm X^{q+}}_{\rm He} Y^{\rm X^{q+}}_{\rm He} \right] {\rm d} s
\end{equation}
The neutral density calculations $n_{\rm N}(\lambda, \beta, R)$ are detailed in \citet{Koutroumpa:2006aa}, based on models by  \citet{Lallement:1984aa} for H and \citet{Dalaudier:1984aa} for He.

The distributions are calculated by following the particle trajectories in the heliosphere, assuming they form a parallel flow with a Maxwellian velocity distribution after crossing the heliopause.
The local neutral density in interplanetary space is calculated based on solar activity and loss processes depending on the solar cycle phase.
Hydrogen trajectories are dominated by the ratio $\mu$ of solar radiation pressure to gravity ($\mu>1$ for the solar maximum, and $<1$ for the solar minimum).
Loss processes for hydrogen consist mainly of resonant charge-exchange with SW protons, and EUV photoionization \citep{QueMerais:2006aa}.
The evolution of these effects with solar activity is monitored and the corresponding changes on the H distribution are modeled based on Ly$\alpha$ back-scattering observations with SOHO/SWAN since 1996 \citep{Koutroumpa:2019aa}.
For helium atoms, radiation pressure is negligible, therefore the trajectories are dominated by gravity and form a localized enhancement downstream from the Sun, the He focusing cone.
Loss processes for He include EUV photoionization \citep{McMullin:2004aa} and electron impact \citep{Rucinski:1989aa}.
Monitoring of the loss processes for He is based on empirical formulas based on CELIAS/SEM data \citep{Bzowski:2013aa}, scaled to \citet{McMullin:2004aa}. 

For the simulations used in this analysis, we calculated the normalized model $Q$ with neutral distributions corresponding to the solar activity phase to each exposures in the filtered \citetalias{Henley:2012aa} sample (omitting high \ion{H}{1} column density sight lines).
The seasonal variation of the normalized model are compared to the observed residuals after subtracting SWCX and the MW contributions in Fig. \ref{yearly_phase}.
The results also show a yearly modulation, since the scaled neutral densities are roughly axisymmetric around the interstellar flow axis, and produce parallax effects for different observer positions (\citealt{Koutroumpa:2012aa}; also see review from \citealt{Kuntz:2019aa}).

\bibliography{ref}{}
\bibliographystyle{aasjournal}

\end{document}